\documentstyle[prd,preprint,aps,epsfig,floats]{revtex}

\begin{document}

\tightenlines
\input epsf.tex
\def\DESepsf(#1 width #2){\epsfxsize=#2 \epsfbox{#1}}
\draft
\thispagestyle{empty}
\preprint{\vbox{ \hbox{UMD-PP-02-50}
\hbox{May 2002}}}

\title{\Large \bf Testing Neutrino Mass Matrices with
Approximate $L_e-L_{\mu}-L_{\tau}$ Symmetry\footnote{This work was
presented at the NUMI Off Axis workshop in Fermilab on May 2, 2002 by
R. N. M.}}

\author{\large H. S. Goh\footnote{e-mail: hsgoh@glue.umd.edu},
 R.N. Mohapatra\footnote{e-mail: rmohapat@physics.umd.edu},
and S.- P. Ng\footnote{e-mail: spng@physics.umd.edu}}

\address{ Department of Physics, University of Maryland\\
College Park, MD 20742, USA}

\maketitle

\thispagestyle{empty}

\begin{abstract}
As neutrino experiments are starting to probe the detailed structure of
the neutrino mass matrix, we present sumrules relating its matrix  
elements for a
class of models with approximate $L_e-L_{\mu}-L_{\tau}$
symmetry and the observables in neutrino oscillation experiments. We show
that regardless of how the
above symmetry is broken (whether in the neutrino sector or the charged
lepton sector), as long as the breaking terms are small, there is a lower
bound on the solar neutrino mixing angle, $sin^22\theta_{\odot}$,
correlated
with the solar mass difference square, $\Delta m^2_{\odot}$, or the mixing parameter, 
$U_{e3}$. We also discuss models where such patterns can arise. 
\end{abstract}

\section{Introduction}
With the evidences for neutrino mass becoming stronger, we have
 reached a stage where it is becoming
possible to contemplate probing the detailed structure of the neutrino
mass matrix. The initial evidences for neutrino mixing from solar and the
atmospheric neutrino data have now been confirmed by several
experiments\cite{atm,solar}. From the recent results of Super-Kamiokande 
and SNO experiments, it now appears that
 the choice for neutrino mixings and therefore the possible profiles for
the neutrino mass matrix within the three neutrino framework have been
considerably
narrowed. There are also evidences for new structures in neutrino masses 
from the LSND\cite{lsnd} and Heidelberg-Moscow neutrinoless double beta
decay experiments\cite{klap}. These latter results need
confirmation. However, if they are included in the neutrino mass analysis,
 they have a profound effect the nature of new physics- for example
confirmation of the LSND results would imply that there are
 four neutrinos rather than three, the fourth one being a
``sterile'' neutrino. Similarly, if the present results of
the Heidelberg-Moscow $\beta\beta_{0\nu}$ are confirmed\cite{klap}, then
 for the first time, we would have at hand
a piece of information that oscillation experiments are unable to
provide, i.e. the overall scale of the neutrino masses. In this paper, we
will take the conservative approach and consider the implications of only
the solar and the atmospheric neutrino results within a three neutrino 
framework. In this case, the mixing pattern is very nearly fixed by the
solar and atmospheric data along with important constraints coming from
CHOOZ-PALO-VERDE experiments\cite{chooz}.

The general Majorana mass matrix for three neutrinos has nine parameters;
three mass eigenvalues, $(m_1, m_2, m_3)$, three mixing angles and three
phases.
Of the three phases, only one is observable in oscillation
expriments. The final goal of neutrino physics is to determine all these parameters,
assuming there are only three neutrinos (as we do in this paper). 

To fit the atmospheric and solar neutrino experiments, we will make the
following choice of mixing angles, which we will incorporate into our
analysis, i.e.  

 (i)  mixing between $\nu_{\mu}$ and $\nu_{\tau}$ is nearly
maximal and 

(ii) there is large mixing between $\nu_e$ and the other active
neutrinos.

\bigskip

 The next piece of information necessary for our analysis is the mass pattern for neutrinos. The data at
the moment does not choose between

(i) hierarchical, i.e. $m_1 \ll m_2 \ll m_3$,

(ii) inverted, i.e. $ m_1\simeq -m_2 \gg m_3$ or 

(iii) degenerate, i.e. $m_1 \simeq m_2 \simeq m_3$ patterns. 

\bigskip
We will choose for our
discussion the inverted pattern since
that corresponds to a very elegant leptonic symmetry,
$L_e-L_{\mu}-L_{\tau}$\cite{models}.
Models with this symmetry have been extensively discussed in
literature\cite{models}. We will here
follow the spirit of the paper\cite{babu}, where two plausible
models for approximate $L_e-L_{\mu}-L_{\tau}$ symmetry were
considered:

(i) where the symmetry was broken by Planck scale effects
and 

(ii) where the charged lepton mass matrix was the source of
leptonic symmetry breaking. 

\bigskip

The first case leads to the so called low solution for the
solar neutrino problem and predicts that KAMLAND experiment currently
running\cite{kamland} should not see a
signal. It also predicts that the parameter $U_{e3}=0$. The
second case, for large $tan\beta$ SUSY models
leads to the LMA solution to the solar neutrino problem and has a
correlation between the $U_{e3}$ and $sin^22\theta_{\odot}$. In this note 
we present a detailed phenomenological analysis of the most general
class of such models.

The motivation for exploring such symmetries of leptonic sector are
manyfold. Historically, symmetries have played a guiding role in the
progress of particle physics as exemplified by symmetries such as isospin,
SU(3),
color, electrweak symmetries etc. Secondly, if the existence of leptonic
symmetries is established, that would make the leptons completely
different from quarks and ideas such as quark-lepton unification would
not be indicated in physics beyond the standard model. This will have
implications for the nature of possible grand unification theories 
e.g. with leptonic symmetries, it will be hard to envision unification
such as SO(10); instead unification schemes based on [SU(3)]$^3$ may be
preferred.

This note focusses on three issues: (A) first, the numerical analysis of
the most general Majorana neutrino mass matrix consistent with weakly
broken
$L_e-L_{\mu}-L_{\tau}$ symmetry in a basis where charged lepton mass
matrix is diagonal and (B) a second case, where the symmetry is exact in
the neutrino mass matrix but broken in 
the charged lepton masses as in ref.\cite{babu} and (C) theoretical
scenarios that make definite predictions for the pattern of
 $L_e-L_{\mu}-L_{\tau}$ breaking in the neutrino mass matrix. We derive
sumrules involving the parameters of the mass matrix and the observables
$sin^22\theta_{\cdot}$, $\Delta m^2_{\odot, A}$ and the mixing parameter
$U_{e3}$ for case (A) and find strong correlation between
these parameters. In particular we find that $sin^22\theta_{\odot}$ in 
this case is necessarily more than $0.95$. This makes it possible to
confront these models
with the next generation neutrino oscillation experiments. We then discuss
case (B) which will be one way to proceed in this framework if the
value of $sin^22\theta_{\odot}$ determined by KAMLAND and
future solar neutrino experiments turns out to be smaller than
$0.95$. In this case, we observe a correlation between $U_{e3}$ and
$sin^22\theta_{\odot}$.  For the specific breaking pattern in the charged
lepton sector described in \cite{babu}, we find a somewhat relaxed lower
bound of $sin^22\theta_{\odot}\geq 0.80$ for $U_{e3}\leq 0.22$. 
 Case (C) presents an example of a model, where
$U_{e3}=0$ is a natural consequence without contradicting any known
results.  We also apply
our sumrules to discuss the effective three neutrino models derived
from 3+1\cite{3p1} models for LSND\cite{lsnd} based on the same symmetry
\cite{rnm}using the seesaw mechanism\cite{seesaw}. 
Depending on whether these models are ruled
out or confirmed, it can provide important information on the direction of
new physics beyond the standard model.

\section{Review of Oscillation phenomenology for three neutrinos}
Before discussing the phenomenology of the model of interest, we repeat a
few basic formulae well known in the literature in order to set the
notation. We express the flavor (or
weak) eigenstates in terms of the mass eigenstates as follows:
\begin{eqnarray}
\nu_\alpha=U_{\alpha i}\nu_i
\end{eqnarray}
where we will use $\alpha$, $\beta$ for the flavor index and $i$, $j $ for
the mass eigenstates index.   $U_{\alpha i}$ are element of the
mixing matrix which diagonalizes the mass matrix of the neutrinos.
The basic formula for the transition probability between two weak
eigenstates is given by
\begin{eqnarray}
|\langle\nu_\beta\vert\nu_\alpha\rangle\vert^2=
\sum_{i,j} A_{\alpha\beta}^{ij}e^{i\frac{\triangle m_{ij}^2}{2p}t}
\label{basic} \cr
A_{\alpha\beta}^{ij}=U_{\alpha i}U_{\alpha j}^*U_{\beta j}U_{\beta i}^*
\label{Afunction}\cr
\end{eqnarray}
$\triangle m_{ij}^2=m_i^2-m_j^2$.
It is easy to see that $(A_{\alpha\beta}^{ij})^*=A_{\alpha\beta}^{ji}= 
A_{\beta\alpha}^{ij}$

Experimental data suggest (i) two distinct mass scales that characterize
the solar and the atmospheric neutrino oscillations: $D3=\triangle
m_{31}^2$ and 
$D2=\triangle m_{21}^2$ with $|D3| >> |D2|$ and (ii) $U_{e3}$ very small.
 In this approximation, the analysis of the two oscillations 
can be separated effectively into $2\times 2$ oscillations. \\
Under the first approximation, the survival probability is, for
atmospheric neutrino,
\begin{eqnarray}
P_{\mu\mu} = 1-4(A_{\mu\mu}^{13}+A_{\mu\mu}^{23})\sin^2(\frac{D_3}{4p}t) 
\end{eqnarray}
for solar neutrino
\begin{eqnarray}
P_{ee} =
1-2(A_{ee}^{13}+A_{ee}^{23})-4(A_{ee}^{12})\sin^2(\frac{D_2}{4p}t) \cr
=(1-2(A_{ee}^{13}+A_{ee}^{23}))(1-\frac{4(A_{ee}^{12})}{1-2(A_{ee}^{13}+
A_{ee}^{23})})\sin^2(\frac{D_2}{4p}t) )
\end{eqnarray}
$A_{ee}^{23}$ and $A_{ee}^{13}$ are both proportional to $U_{e3}^2$ and so the 
oscillation amplitude of atmospheric $\nu_e$ is suppressed. 
Note that the analysis becomes effectively that of two neutrino 
oscillation. From the above equation, we can read out the mixing angle
\begin{eqnarray}
\sin^22\theta_{A}=4(A_{\mu\mu}^{13}+A_{\mu\mu}^{23})=4U_{\mu
3}^2(1-U_{\mu 3}^2)\cr
\sin^22\theta_{\odot}=\frac{4(A_{ee}^{12})}{1-2(A_{ee}^{13}+
A_{ee}^{23})}=\frac{4U_{e1}^2U_{e2}^2}{1-2U_{e3}^2(1-U_{e3}^2)}
\end{eqnarray}
Note that the above formulae for the survival probabilities remain valid
even in the presence of complex phases in the mixing matrices.
We will use these formulae in deriving the effective mixing
angles for the atmospheric and solar neutrino oscillations\cite{reviews}.

\section{Approximate $L_e-L_{\mu}-L_{\tau}$ symmetry for leptons} 
In order to explore the possible existence of symmetries in the lepton
sector of physics beyond the standard model, we consider two cases in this
section. In the first one we allow the most general small symmetry
breaking terms in the neutrino mass mass matrix keeping the charged lepton
sector completely symmetry conserving and in the second case, we keep the
neutrino sector symmetry conserving and allow small breakings in the
charged lepton sector. In the subsequent section we discuss physics
scenarios where these two cases can arise.

\subsection{Symmetry breaking in the neutrino sector and sumrules for mass
matrix elements}
 In this section, we will consider 
 the neutrino mass matrix for the case with weakly broken
$L_e-L_{\mu}-L_{\tau}$ symmetry in the neutrino sector. We will assume
the dominant terms to respect $L_e-L_{\mu}-L_{\tau}$ symmetry with 
the symmetry breaking terms to be less than one third of the 
symmetry preserving ones and work in the perturbative approximation
of these parameters. We find sumrules that relate the symmetry breaking 
parameters to the observables in neutrino oscillation experiments. 
One can then use these to test for the possible existence of the leptonic
symmetry $L_e-L_{\mu}-L_{\tau}$.

To proceed with the derivation of the sumrules, we write down the neutrino
mass matrix in the exact symmetry limit i. e.
\begin{eqnarray}
M^{(0)}_\nu~=~m\left(\begin{array}{ccc} 0 &
\sin \theta & \cos\theta\\ \sin\theta & 0 & 0\\ \cos\theta & 0 &
0\end{array}\right).
\end{eqnarray}
We add to it
the symmetry breaking matrix of the following general form: 
\begin{eqnarray}
\triangle M_\nu=m~\left(\begin{array}{ccc} z &
0 & 0\\ 0 & y & d\\ 0 & d & x\end{array}\right).
\end{eqnarray}
The full neutrino mass matrix is $M_{\nu}~=~M^{(0)}_{\nu}+\triangle
M_{\nu}$.
The charged lepton mass matrix can be cast into a diagonal form in this
basis in the symmetry limit by redefining the parameters $m$ and angle
$\theta$. In
this example we do not assume any symmetry
breaking terms in the charged leptn sector.

 In the perturbative
approximation, we find the following sumrules
involving the neutrino observables and the elements of the neutrino mass
matrix. The two obvious relations are
 \begin{eqnarray}
\sin^22\theta_{A}~=~\sin^22\theta + O(\delta^2)\cr
D_3 ~=~ \triangle m_A^2~=~-m^2 + 2 \triangle m^2_{\odot}~+~ O(\delta^2)\cr
\end{eqnarray}
The nontrivial relations that also hold for this model are:
\begin{eqnarray}
\sin^22\theta_{\odot}~=~1-(\frac{\triangle m_\odot^2}{4\triangle
m_A^2}-z)^2~+~O(\delta^3) \cr
\frac{D_2}{D_3}\equiv \frac{\triangle m_\odot^2}{\triangle
m_A^2}~=~2(z+\vec{v}\cdot\vec{x})~+~O(\delta^2)\cr
U_{e3}~=~\vec{A}\cdot(\vec{v}\times\vec{x})~+~O(\delta^3)\cr
\end{eqnarray} 
where $\vec{v}=(\cos^2\theta,\sin^2\theta,\sqrt{2}\sin\theta\cos\theta)$, 
$\vec{x}=(x,y,\sqrt{2}d)$ and
$\vec{A}~=~\frac{1}{\sqrt{2}}(1,1,0)$. $\delta$
in Eq. (8) and (9) represents the 
small parameters in the mass matrix. These equations represent one of the
main results of this paper. Below we study their implications. Finally,
there is the relation $\langle m \rangle_{\beta\beta}~=~ z$. This is an
exact relation with no $O(\delta^n)$ corrections. Note that in
the first of the two equations in Eq. (9), if we set $z=0$, we get a
relation which is similar to the one that holds in the Zee
model\cite{koide} but apparently valid for a wider class of models.

Experimental data requires $\frac{\triangle m_\odot^2}{\triangle m_A^2}<
0.1$, which implies that our model predicts
$\sin^22\theta_{\odot} > 0.95$ for $|z|\leq 0.2$. Combining
this with the mean value for the $\Delta m^2_{Atmos}$, we predict that
the effective neutrino mass in the neutrinoless double beta decay in this
case is given by $\langle m \rangle_{\beta\beta} \simeq 0.009$ eV. The
detailed correlation between $sin^22\theta_{\odot}$ and $\Delta
m^2_{\odot}$ is given in Fig. 1. It includes also the second order
perturbation effects not present in Eq.(9). In Fig. 2, we present the
predictions for
$U_{e3}$ in this model. We see that $U_{e3}$ is pretty much allowed up to
its present limit. Therefore, this model can be tested once the KAMLAND
experiment is complete assuming that it confirms the LMA solution to the
solar neutrino problem.

\begin{figure}[h!]
\begin{center}
\epsfxsize15cm\epsffile{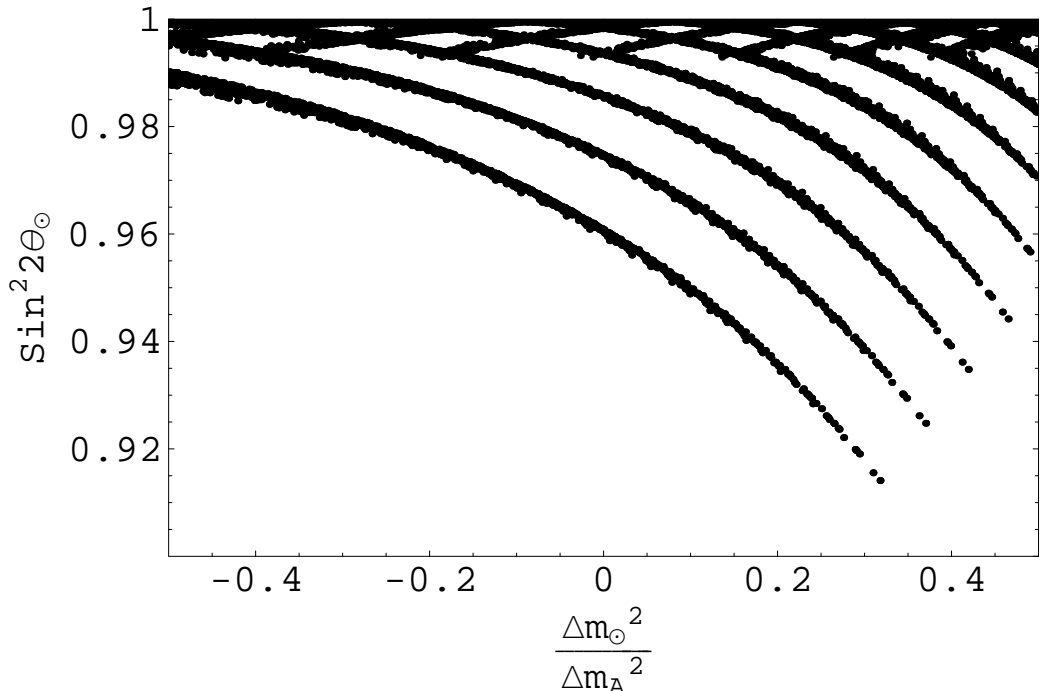}
\caption{ 
The figure shows the predictions for $sin^22\theta_{\odot}$ as
a function $\Delta m^2_{\odot}/\Delta m^2_{A}$ for different values
for the symmetry breaking parameters. The left most line corresponds to
$z=-0.2$ and the right most (only partially visible) to $z=0.16$. The thickness of the individual
lines reflect the higher order contributions.
\label{fig:cstr1}}
\end{center}
\end{figure}

\begin{figure}[h!]
\begin{center}
\epsfxsize15cm\epsffile{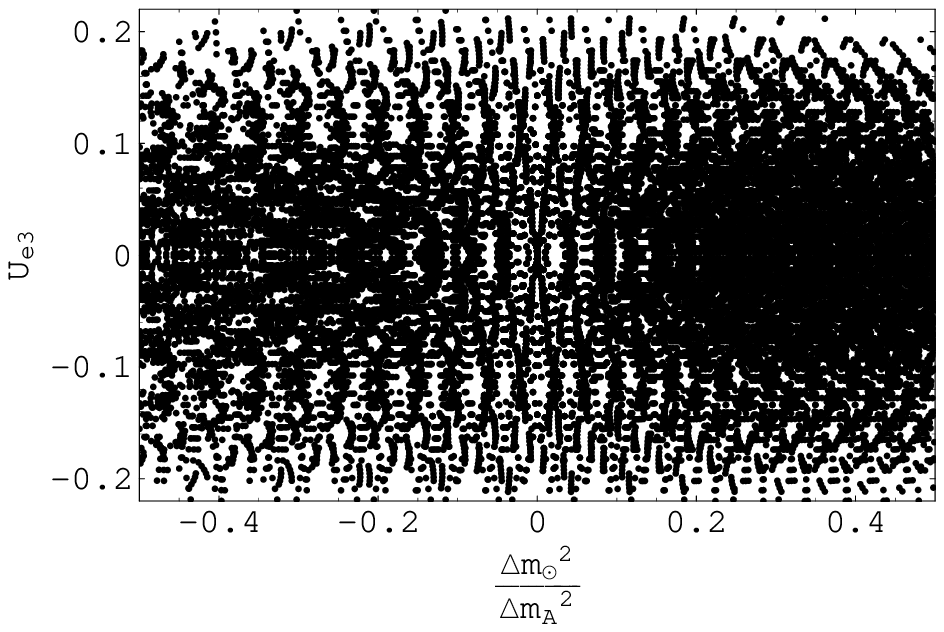}
\caption{ This figure shows the predictions for $U_{e3}$ as a function of
$\Delta m^2_{\odot}/\Delta m^2_{A}$. Note that practically all values
of $U_{e3}$ are allowed regardless of the value of  $\Delta
m^2_{\odot}/\Delta m^2_{A}$
 \label{fig:cstr1}} \end{center}
\end{figure}

An immediate implication of our result above is that
if $sin^22\theta_{\odot}$ is found to be smaller than the above value
$0.95$, then either $L_e-L_\mu-L_\tau$ symmetry has to be very badly
broken or there must be contributions to symmetry breaking from the
charged lepton sector.

\subsection{Effect of symmetry breaking in the charged lepton sector} 
 We now include the charged lepton sector and following the
suggestion of Ref.\cite{babu}, consider the charged lepton
mass matrix of the form
\begin{eqnarray}
M_e~=~\left(\begin{array}{ccc} 0 &
0 & a\\ 0 & b & 0\\ a' & 0 & 1\end{array}\right) \frac{m_\tau}{q} \cr
q=\sqrt{1+a^2+a'^2}
\end{eqnarray}
We work in the charged lepton mass eigenstate basis. In this basis, the
neutrino mass matrix can be rotated from its 
original mass matrix by $V_L$
\begin{eqnarray}
M_\nu \longrightarrow V_L^\dagger M_\nu V_L
\end{eqnarray}
where $V_L$ is defined by $V_L^\dagger(M_eM_e^\dagger)V_L$ 
is diagonal. The charged lepton masses require
\begin{eqnarray}
\frac{a^2a'^2}{q^2}\simeq (\frac{m_e}{m_\tau})^2\simeq 7.7\times 10^{-8}
\cr
\frac{b^2}{q^2}\simeq (\frac{m_\mu}{m_\tau})^2\simeq3.4\times 10^{-3}
\end{eqnarray}
we need $a$ to be relatively big to be able to generate the large
mixing angle solution to the solar neutrino problem; $a'$ is then has to
be small and we approximate it to be zero.  
In the charged lepton mass eigenstate basis, the neutrino mass matrix is
found to be
\begin{eqnarray}
M_\nu ~=~\left(\begin{array}{ccc} -\frac{2a\cos\theta}{1+a^2} &
\frac{\sin\theta}{\sqrt{1+a^2}} & \frac{(1-a^2)\cos\theta}{1+a^2}\\
\frac{\sin\theta}{\sqrt{1+a^2}} & 
0 & \frac{a\sin\theta}{\sqrt{1+a^2}} \\ \frac{(1-a^2)\cos\theta}{1+a^2} &
\frac{a\sin\theta}{\sqrt{1+a^2}} & 
\frac{2a\cos\theta}{1+a^2}\end{array}\right)m 
\end{eqnarray}
To this mass matrix, we must add the renormalization group corrections due
to the charged lepton mass, which then induces the solar mass splitting.
As was noted in \cite{babu}, this process of rotation of the charged
lepton and the renormalization group corrections
lead to a nonzero value for the $U_{e3}$ parameter correlated with the
solar mixing angle, $sin^22\theta_{\odot}$. In Fig. 3, we show the correlation between these two
parameters. We note that as $U_{e3}$ reduces, the value of
$sin^22\theta_{\odot}$ increases. For $U_{e3}\leq 0.22$, we get
$sin^22\theta_{\odot} \geq 0.85$. This can therefore be used to test the
model. It is important to note that the solar mixing angle
$sin^22\theta_{\odot}$ will be more precisely measured by KAMLAND (down to
the level of $0.91$) if the
LMA solution to the solar neutrino puzzle is correct. There are many
experimental projects (such the off axis NUMI beam\cite{para}, JHF and the
neutrino
factory proposals) whose main goal is to measure (or limit) $U_{e3}$ down
to the level of a few percent. Clearly this will then very severely
constrain or rule out the case where the $L_e-L_{\mu}-L_{\tau}$ symmetry
is broken in the charged lepton sector.

\begin{figure}[h!]
\begin{center}
\epsfxsize15cm\epsffile{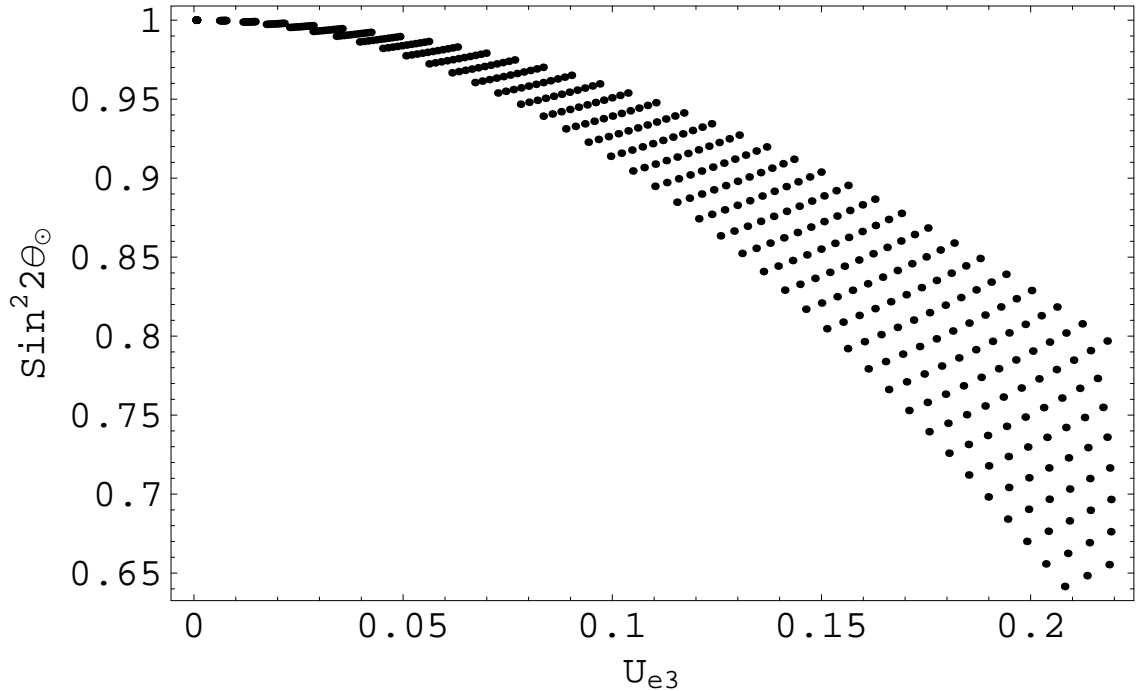}
\caption{This figure shows the correlation between $U_{e3}$ and
$sin^22\theta_{\odot}$ for the model where $L_e-L_{\mu}-L_{\tau}$ symmetry
is broken by the charged lepton sector. Note that present upper limit of
$0.22$ corresponds to a minimum value of $sin^22\theta_{\odot}\geq 0.80$
and $U_{e3}\leq 0.16$ corresponds to $sin^22\theta_{\odot}\geq 0.90$.
The different lines in the figure correspond to various values for the 
$sin^22\theta_A$ with the rightmost line corresponding to the maximal
value of 1. The lines in this figure correspond to the following relation
between $U_{e3}$ and $sin^22\theta_{\odot}$ 
i.e. $sin^22\theta_{\odot}~=
~\left[1-\frac{1+cos^2\theta_A}{sin^2\theta_A}U^2_{e3}\right]^2/
\left(1-U^2_{e3}\right)^2$.
 \label{fig:cstr1}} \end{center}
\end{figure}

\section{Theoretical origin of the mass matrices}

In this section, we will discuss possible theoretical origin of the mass
matrices analyzed above in simple gauge models. We will first
present a simple extension of the standard model by adding two
additional singlet right handed neutrinos, $N_1, N_2$ assigning
them $L_e-L_\mu-L_\tau$ quantum numbers of $=1$ and $-1$
respectively. Denoting the standard model lepton doublets by
$\psi_{e,\mu,\tau}$, the $L_e-L_\mu-L_\tau$ symmetry allows the
following new couplings to the Lagrangian of the standard model:
\begin{eqnarray}
{\cal L}'~=~(a\bar{\psi_{\tau}} + b\bar{\psi}_{\mu})H N_2 +
\bar{\psi}_eHN_1 + M N^T_1C^{-1}N_2
\end{eqnarray}
where $H$ is the Higgs doublet of the standard model; $C^{-1}$ is the
Dirac charge conjugation matrix.
We add to it the symmetry breaking mass terms for the right handed
neutrinos, which are soft terms, i.e.
\begin{eqnarray}
{\cal L}_B~=~ \epsilon(M_1 N^T_1C^{-1}N_1 + N^T_2C^{-1}N_2) + H.c.
\end{eqnarray}
with $\epsilon \ll 1$. These terms break  $L_e-L_\mu-L_\tau$ by two units
but since they are dimension 3 terms, they are soft and do not induce any
new terms into the theory.

It is clear from the resulting mass matrix for the $\nu_L, N$ system 
that the linear combination $b\nu_{\mu}-a \nu_e)$ is massless and the
atmospheric oscillation angle is given by $tan \theta_A = b/a$; for $a\sim
b$, the $\theta_A$ is maximal. The seesaw mass matrix then takes the
following form (in the basis $(\nu_e, \tilde{\nu}_{\mu}, N_1, N_2)$ with 
$\tilde{\nu}_{\mu}\equiv a \nu_{\mu} + b\nu_{\tau}$):
\begin{eqnarray}
M~=~\left(\begin{array}{cccc} 0 & 0 & m_1 & 0\\0 & 0 & 0 & m_2\\m_1 & 0 &
\epsilon M_1 & M \\ 0 & m_2 & M & \epsilon M_2\end{array}\right)
\end{eqnarray}
The diagonalization of this mass matrix leads to the mass matrix of the
form discussed in Eq. (6) and (7) with the constraint that $d/x =
y/d= m_2/m_1$. This
has the implication that $U_{e3}=0$. One could get nonzero values for
$U_{e3}$ by further extending the model to include another heavy right
handed neutrino.

\section{Symmetry breaking induced by sterile neutrino}

In this section, we consider an application of our sumrules to 
 a four neutrino model which
accomodates the LSND results within the 3+1 scenario\cite{3p1}
and where the mass matrix in the three active neutrino subsector
obeys $L_e-L_{\mu}-L_{\tau}$ symmetry. The only symmetry breaking terms in
this case are due to the mixing of the active neutrinos with the sterile
one. The mass matrix in this case is given by:
\begin{eqnarray}
M^{(4)}_{\nu}~=~\left(\begin{array}{cccc} 0 & m_1 & m_2 & \delta_1 \\
m_1 & 0 & 0 & \delta_2 \\
m_2 & 0 & 0 & \delta_3 \\
\delta_1 & \delta_2 & \delta_3 & \Delta \end{array}\right)
\end{eqnarray}
Such patterns are predicted by minimal quark lepton symmetric seesaw 
models for the sterile neutrino as discussed in \cite{rnm}.
In this mass matrix $\delta_i \ll \Delta$; The diagonalization of this
mass matrix can be carried by using the seesaw technique\cite{balaji} 
which leads to
the effective $3\times 3$ active neutrino submatrix of the form that now
has induced $L_e-L_{\mu}-L_{\tau}$ breaking terms.
 \begin{eqnarray}
M^{(3)}_{\nu}~=~ \left(\begin{array}{ccc} -\frac{\delta^2_1}{\Delta} &
m'_1
& m'_2 \\ m'_1 & -\frac{\delta^2_2}{\Delta} &
-\frac{\delta_2\delta_3}{\Delta}
\\ m'_2 & - \frac{\delta_2\delta_3}{\Delta} & -\frac{\delta^2_3}{\Delta}
\end{array}\right)
\end{eqnarray}
where $m'_{1,2}$ has new small contributions from the diagonalization
(terms of the form $\frac{\delta_i\delta_j}{\Delta}$).
In order to fit the LSND results, we must have
$(\delta_1\delta_2/\Delta^2)\simeq 0.03$. For $\Delta \simeq 1 eV$, this
implies that roughly $\delta_i\simeq 0.17$. Using this along with the
sumrules of section
2, we immediately conclude that the model predicts $\Delta
m^2_{\odot}\simeq 10^{-3}$ eV$^2$, which is much too large a value. This
simple version of the model is therefore in contradiction with data.
If we want to fit the solar neutrino data in the context of such a model,
then it would predict a maximum value for the $\nu_{\mu}-\nu_{e}$
transition probability at the level of $10^{-4}$. Unfortunately this is
below the sensitivity of Mini-Boone\cite{louis} for detecting this
process.

\section{Conclusion}
   In summary, we have outlined the predictions of the assumption that
the neutrino mass matrix obeys a softly broken $L_e-L_{\mu}-L_{\tau}$
symmetry either in the neutrino or in the charged lepton sector. These
predictions if confirmed will constitute evidence both for an inverted
mass hierarchy among neutrinos as well as for the existence of an
approximate $L_e-L_{\mu}-L_{\tau}$ symmetry, a new symmetry of physics
beyond the standard model. It is interesting that these predictions can be
tested in the current round of solar neutrino experiments such as KAMLAND
and more definitively once future neutrino projects are able to improve
the limits on the parameter $U_{e3}$. The presence or absence of such
leptonic symmetries will clearly be of fundamental significance for
physics beyond the standard model.

\bigskip

\bigskip

\bigskip

 \section*{Acknowledgements}

   One of the authors (RNM) is grateful to the Institute for Nuclear
Theory at the University of Washington for hospitality when part of the
work was completed. He would also like to thank K. S. Babu and
B. Kayser for discussions. This work is supported by the National Science
Foundation Grant No. PHY-0099544.

Note added: After this paper was posted in the Archives, it was brought to
our attention that the model presented in section (iv) of 
this paper was also discussed by W. Grimus and L. Lavoura in the paper now
cited in Ref.\cite{models}. It was also brought to our attention that
mass matrices with inverted hierarchy were analyzed in \cite{king}.


\begin{thebibliography}{99}

\bibitem{atm} SuperKamiokande collaboration, Y. Fukuda et. al.,
{\it Phys. Rev. Lett.} {\bf 82}, 2644 (1999).

\bibitem{solar} B.T. Cleveland et. al., {\it Astrophys. J.} {\bf 496}, 505
(1998); R. Davis, {\it Prog. Part. Nucl. Phys.} {\bf 32}, 13 (1994);
SuperKamiokande collaboration, Y. Fukuda et. al.,
{\it Phys. Rev. Lett.} {\bf 81}, 1158 (1998); Erratum $ibid$., {\bf 81},
4279 (1998) and $ibid$., {\bf 82}, 1810 (1999); Y. Suzuki, {\it Nucl.
Phys. Proc. Suppl.} {\bf B 77},  35 (1999); Y. Fukuda et. al.,
hep-ex/0103032; SAGE collaboration, J.N. Abdurashitov et. al.,
{\it Phys. Rev.} {\bf C 60}, 055801 (1991); GALLEX collaboration,
W. Hampel et. al.
{\it Phys. Lett.} {\bf B447}, 127 (1999); SNO collaboration,
Q.R. Ahmed et. al., {\it Phys. Rev. Lett.} {\bf 87}, 071301 (2001);
for SNO neutral current data, see www.sno.queensu.ca.

\bibitem{lsnd} C. Athanassopoulos {\it et al.,}
Phys. Rev. {\bf C54} (1996) 2685;
C. Athanassopoulos {\it et al.,}, Phys. Rev. {\bf C58}, 2489 
 (1998).

\bibitem{klap} H. Klapdor-Kleingrothaus et al., Mod. Phys. Lett. {\bf
A16}, 2409 (2002).

\bibitem{chooz} CHOOZ collaboration, M. Apollonio et. al., {\it
Phys. Lett.} {\bf B466}, 415 (1999); Palo-verde collaboration: F. Boehm et
al., Phys. Rev. Lett. {\bf 84}, 3764 (2000).

\bibitem{models}  S. T. Petcov, Phys. Lett. {\bf 110
B},245 (1982);
R. Barbieri, L. Hall, A. Strumia and N. Weiner, JHEP 9812,
017  (1998);
A. Joshipura and S. Rindani, Eur.Phys.J. {\bf C14}, 85
(2000);
R. N. Mohapatra, A. Perez-Lorenzana, C. A. de S. Pires,
Phys. Lett. {\bf B474}, 355 (2000);
T. Kitabayashi and M. Yasue,
Phys. Rev. {\bf D 63}, 095002 (2001); Phys. Lett. {\bf B 508}, 85 (2001); 
hep-ph/0110303;  L. Lavoura, Phys. Rev. D 62, 093011 (2000);
 W. Grimus and L. Lavoura, Phys. Rev. D 62, 093012 (2000);
 J. High Energy Phys. 09, 007 (2000); J. High Energy Phys. 07, 045 (2001);
R. N. Mohapatra, hep-ph/ 0107274; 
A. Aranda, C. Carone and P. Meade,
hep-ph/0109120;
 K. S. Babu and R. N. Mohapatra, hep-ph/0201176;  Phys. Lett. B (to
appear); Duane A. Dicus, Hong-Jian He, John N. Ng, Phys. Lett. {\bf B
536}, 83 (2002).

\bibitem{babu} See K. S. Babu and R. N. Mohapatra, Ref.\cite{models}.

\bibitem{kamland} KAMLAND collaboration, A. Piepke, {\it Proceedings of
the Neutrino2000 conference}, Nucl. Phys. Proc. Suppl. {\bf B 91}, 99
(2001).

\bibitem{3p1} S. M. Bilenky, C. Giunti, W. Grmus and
T. Schwetz,  Phys. Rev. {\bf D60}, 073007 (1999); 
V. Barger, B. Kayser, J. Learned, T. Weiler, K. Whisnant, Phys. Lett.
{\bf B489}, 345 (2000).


\bibitem{rnm} R. N. Mohapatra, hep-ph/0107274; Phys. Rev. {\bf D 64 },
091301 (2001). 

 \bibitem{seesaw}  M. Gell-Mann, P. Ramond and R. Slansky, in {\it
Supergravity}, eds. P. van Niewenhuizen and D.Z. Freedman (North
Holland 1979); T. Yanagida, in Proceedings of {\it Workshop on
 Unified Theory and Baryon number in the Universe}, eds.
O. Sawada and A. Sugamoto (KEK 1979);  R.N. Mohapatra and
G. Senjanovi{\'c}, Phys. Rev. Lett. {\bf 44}, 912 (1980).

\bibitem{reviews} For reviews and other references, see E. Kh. Akhmedov,
Lectures in Trieste Summer School, (1999), ed. G. Senjanovi\'c and
A. Smirnov (World Scientific); C. Gozales-Garcia and Y. Nir,
hep-ph/0202058.

\bibitem{koide} Y. Koide, Phys. Rev. {\bf D 64}, 077301 (2001);
P. Frampton and S. L. Glashow, Phys. Lett. {\bf B 461}, 95 (1999).

\bibitem{para} A. Para, hep-ex/0110032.

\bibitem{balaji}  K. Balaji, A. Perez-Lorenzana and A. Smirnov,
hep-ph/0101005; A. Perez-Lorenzana and Carlos A. de S. Peres, Phys. Lett.
{\bf B 522}, 297 (2001). 


\bibitem{louis} W. Louis, Invited talk at the "INT Miniworkshop on
Neutrino Masses and Mixings", Seattle, Washington, April, 2002.

\bibitem{king} S. F. King. hep-ph/0204360.

\end{thebibliography}
\end{document}